\pdfoutput=1
\documentclass{article}
\RequirePackage[numbers, authoryear]{natbib}
\usepackage{tablefootnote} 
\usepackage{threeparttable}
\usepackage{comment}
\usepackage{draftcopy}
\usepackage{graphicx}%
\usepackage{multirow}%
\usepackage{amsmath,amssymb,amsfonts}%
\usepackage{amsthm}%
\usepackage{mathrsfs}%
\usepackage[title]{appendix}%
\usepackage{xcolor}%
\usepackage{textcomp}%
\usepackage{manyfoot}%
\usepackage{booktabs}%
\usepackage{algorithm}%
\usepackage{algorithmicx}%
\usepackage{algpseudocode}%
\usepackage{listings}%
\usepackage{enumerate}
\usepackage[colorlinks=true, citecolor=black, urlcolor=blue]{hyperref}
\usepackage{authblk}

\providecommand{\keywords}[1]
{
  \small	
  \textbf{\textit{Keywords---}} #1
}

\begin{document}

\title{Semiparametric Growth-Curve Modeling in Hierarchical, Longitudinal Studies}
\author{Rajesh Selukar}
\affil{SAS Institute Inc., Cary NC \\rajesh.selukar@sas.com} 
\date{}

\maketitle

\begin{abstract}
Modeling of growth (or decay) curves arises in many fields such as microbiology, epidemiology, marketing, 
and econometrics. Parametric forms like Logistic and Gompertz are often used for modeling such monotonic patterns. 
While useful for compact description, the real-life growth curves rarely follow these parametric forms perfectly.  Therefore, the 
curve estimation methods that strike a balance between prior information in the parametric form and fidelity with the 
observed data are preferred.  In hierarchical, longitudinal studies the interest lies in comparing the growth curves of 
different groups while accounting for the differences between the within-group subjects.  This article describes a 
flexible state space modeling framework that enables semiparametric growth curve modeling for the data generated from 
hierarchical, longitudinal studies.  
The methodology, a type of functional mixed effects modeling, is illustrated with a real-life example of bacterial 
growth in different settings.  
\end{abstract}


\section{Introduction}
Study of growth and decay patterns arises in many fields.  For example, microbiologists study 
the growth of bacteria in different settings, epidemiologists study the increase in the number 
of infected patients after a disease outbreak and, chemists study the decay of the active 
agents in chemical compounds to determine their shelf lives.  
The analysis of experimental or observational data from such studies is used for a variety 
of purposes, such as, validating the understanding of the inner workings of the process, 
comparing the growth/decay patterns in different settings, extrapolating the 
growth patterns in the future, and so on.  

An important step in such analysis is the estimation of the underlying growth (decay) curves.  
Parametric forms like Logistic and Gompertz, as well as nonparametric forms such as smoothing splines, are often 
used for modeling such monotonic patterns. 
The usefulness and popularity of parametric approach is clear from the large number of R packages available in the 
Comprehensive R Archive Network (CRAN), which support parametric growth-curve modeling.   
A good summary of the state of art in the parametric growth-curve modeling is provided in \cite{bioGrow1}, 
which describes the functionality of an R package, \emph{biogrowth}.  
In nonparametric growth curve modeling, smoothing splines with and without monotonicity constraints
is a popular approach, e.g., see, \cite{MonoSpline1, MonoSpline2} and the references therein.
In \cite{Covid1}, the spread of COVID-19 cases is modeled using
a state space modeling based approach.  They, however, model 
the growth rate (derivative of the growth curve) rather
than the growth curve itself.
All these references, with the exception
of \cite{bioGrow1}, are primarily concerned with the modeling of a single growth curve. 
In hierarchical, longitudinal studies, multiple growth curves are
generated as a result of monitoring multiple subjects from a group (or multiple groups).  
In such cases the interest lies in estimating the group-mean-curves, after accounting
for the subject specific differences.  For these types of problems the dominant
approaches have been based on linear mixed-effects, nonlinear mixed-effects, and functional mixed-effects
modeling, e.g., see \cite{grimm, blozis}.  In these references, when  linear mixed-effects or functional mixed-effects
models are used, the group-mean-curves are often linear or polynomial
smoothing splines, which, while flexible, do not use the prior monotonicity information of the growth curves.
In nonlinear mixed-effects modeling, it is assumed that the subject curves are generated
by parametric growth curve models such as Logistic and some or all of the model parameters, like
the growth rate, for each subject are assumed to be random perturbations of their
mean population values.  However, in this case, the subject curves being fully parametric can be a
restrictive assumption. 

In this article, the analysis of multiple growth curves generated by hierarchical, longitudinal studies
is done using functional mixed-effects (FME) modeling.  An FME model 
for growth curves decomposes a subject specific curve as follows:
\begin{eqnarray}
\text{
subject-curve} & = & \text{group-mean-curve + regression-effects +} \nonumber \\
                          & &       \text{deviation-from-mean-curve + noise}  \nonumber
\end{eqnarray}
The group-mean-curve, which is expected to follow a growth curve pattern, is called a 
functional fixed-effect, the deviation-from-mean-curve
is called a functional random-effect, regression-effects can either be fixed or random,
and the noise is a zero-mean, Gaussian error term. In this article, 
the group-mean-curve is modeled semiparametrically according to a simple
modification of the
state space model based approach described in section 3 of \cite{ans1}. 
Our modification makes their approach computationally feasible in much
wider settings.
Making their elegant growth curve modeling approach
easily accessible to the wider applied research community is one
of the goals of this article.
All the FME models considered in this article are formulated as linear, Gaussian, state space models, 
SSMs for short.  This provides many additional advantages
such as modeling flexibility, computational efficiency, and easy interpretability (see
 \cite{Guo1, selu_r:15}).  
The analysis
for the illustrations in this article is carried out by using the CSSM procedure in the SAS Viya/Econometrics 
software, see \cite{cssm1}.  The data and the SAS code used for these illustrations is available on request.

  The remainder of the article is organized as follows: Section~\ref{formulation}
lays out the required SSM framework (the parametric growth curve formulation is in 
Section~\ref{form1}, the semiparametric formulation is in Section~\ref{form2}, and 
the FME model formulation is in Section~\ref{form3}), Section~\ref{example} contains an
example that explains the methodology, and finally Section~\ref{final} concludes.
\section{State Space Modeling Framework}
\label{formulation}
SSMs have long been used to model
a variety of sequential data, such as time series,
panels of time series, and longitudinal data.  
Many different forms of SSMs are used in practice.
In order to handle the formulation of FME models for the
longitudinal data, this article uses an SSM form that is similar to the one described
in chap. 6, sec. 4 of \cite{dk}.   The notation and additional details are included in the appendix, Appendix~\ref{SSMFramework}.

In this article when vectors and matrices are described in inline mode, they are written row-wise with
rows separated by semicolons.
\subsection{SSM Formulation of a Parametric Growth Curve}
\label{form1}
Let $y_{t}$ denote noisy measurements on a growth/decay curve $f(\boldsymbol{\theta}, t)$.  That is,
\begin{equation}
y_{t} =  f(\boldsymbol{\theta}, t) + \epsilon_{t} \label{eq1}
\end{equation}
where $\epsilon_{t}$ is a noise sequence of independent, zero-mean, Gaussian variables with variance 
$\sigma^{2}_{\epsilon}$,
and  the latent curve, $f(\boldsymbol{\theta}, t) = constant + scale \; g(\boldsymbol{\theta}, t)$, 
is a smooth, monotonic
function of $t$, which might depend on some parameter vector $\boldsymbol{\theta}$;
$constant$ and $scale$ are auxiliary parameters that do not depend on $\boldsymbol{\theta}$.
A few polpular choices of $g(\boldsymbol{\theta}, t)$ are shown in Table~\ref{GDCurves};
several more examples can be found in \cite{zw}.
\begin{table*}[t]
\begin{center}
\caption{Examples of $g(\boldsymbol{\theta}, t)$ in Popular Growth/Decay Curves}
\label{GDCurves}
\begin{tabular}{@{}l  c c c c @{}}
\hline
\multicolumn{1}{l}{Name} &  & & \multicolumn{1}{c}{$g(\boldsymbol{\theta}, t)$}  & 
\multicolumn{1}{c}{Parameter Vector  $\boldsymbol{\theta}$}\\
\hline
Linear    & & & $t$  & \\
\\
Exponential    & & &  $\exp(-\rho \;t)$ & $\rho $\\
\\
Logistic & & & $\frac{1}{(1 + \phi \exp(-\rho \;t))}$ & $\phi, \rho $ \\
\\
Gompertz &  & & $\exp(-\phi \exp(-\rho \;t))$& $ \phi, \rho $ \\
\\
Richards & & & $(1 + \phi \exp(-\rho \;t))^{(-1/\nu)}$& $ \phi, \rho, \nu $  \\
\hline
\end{tabular}
\end{center}
\end{table*}
When the functional form of the latent curve, $f(\boldsymbol{\theta}, t)$, is completely known,
the unknown parameters, $(constant, scale, \boldsymbol{\theta}, \sigma^{2}_{\epsilon})$, can be estimated by
nonlinear least-squares. Alternatively, the estimation of $f(\boldsymbol{\theta}, t)$
can be accomplished by using the SSM form of Equation~\ref{eq1},
which has many additional advantages.  The SSM form is based on a simple two-dimensional recursion.
Suppressing $\boldsymbol{\theta}$ for notational simplicity, $f(t+h) = constant + scale \; g(t+h)$ can be
expressed in terms of $f(t)$ as
\[
\begin{bmatrix}
constant + scale \; g(t+h) \\
scale
\end{bmatrix}
  = 
\begin{bmatrix}
1 & (g(t+h)  - g(t))\\
0 & 1
\end{bmatrix}
\begin{bmatrix}
constant + scale \; g(t)   \\
scale
\end{bmatrix} 
\]
This recursion leads to the formulation of Equation~\ref{eq1} as
an SSM with two dimensional latent state vector $\boldsymbol{\alpha}_{t}$:
\begin{eqnarray}
y_{t} & =  & [1 \; 0]  \; \boldsymbol{\alpha}_{t} + \epsilon_{t}   \; \quad \text{Observation Equation} \nonumber \\
\boldsymbol{\alpha}_{t+h} & = & \mathbf{T}_{t}^{t+h} \;\boldsymbol{\alpha}_{t}  \quad  
            \quad \quad \text{State Equation}  \label{BasicSSM}\\
\boldsymbol{\alpha}_{0} & = &  \boldsymbol{\delta} \quad  \quad\quad \quad\quad\quad
\text{Diffuse Initial Condition} \nonumber
\end{eqnarray}
where $\boldsymbol{\alpha}_{t} = [f(t); \; scale]$, the transition matrix $\mathbf{T}_{t}^{t+h} = [1 \; (g(t+h)  - g(t)); \; 0 \;1]$ 
and, the initial condition is a two-dimensional latent vector, $\boldsymbol{\delta}$.  
Note that the state equation of this SSM has no disturbance term.  Standard (diffuse) Kalman filtering enables
the maximum (marginal) likelihood estimation of $\boldsymbol{\theta}$ and $\sigma^{2}_{\epsilon}$. 
Once $\boldsymbol{\theta}$ and $\sigma^{2}_{\epsilon}$ are estimated, (diffuse) Kalman smoothing
enables the estimation of all latent quatities such as the initial condtion $\boldsymbol{\delta}$, and 
$\boldsymbol{\alpha}_{t} = [f(t); \; scale]$
at all time points $t$.  

\begin{figure}[t]
\begin{center}
\includegraphics[scale=0.50]{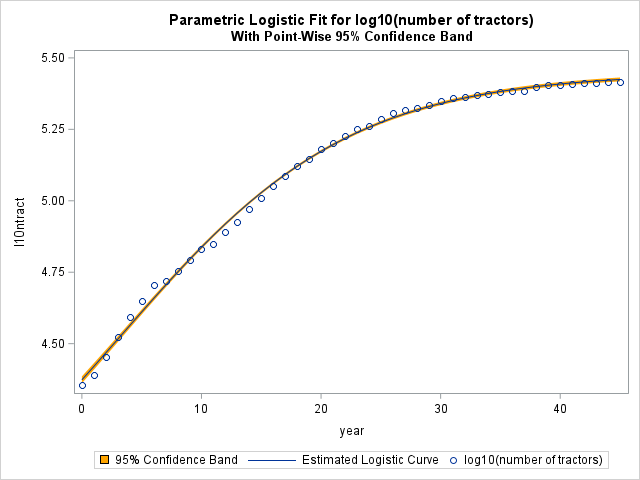}
\caption{Parametric Logistic Fit Based on the SSM Formulation}
\label{logCurve}
\end{center}
\end{figure}

As an illustration, consider the \emph{greek\_tractors} data set considered in \cite{bioGrow1}.  It contains
the number of tractors  registered in Greece in a 45 year period between 1961 to 2006.  They fit the
Logistic curve to $y = \log_{10}(\text{number of tractors})$, by nonlinear least squares.  After substituting the parameter
estimates, their fitted Logistic curve has the following expression:
\begin{equation}
f(t) = 3.605 + 1.844/(1 + 1.398*exp(- 0.104*t)) \label{logEq}
\end{equation}
where the time $t$, which starts at 0, measures years since 1961.  As it turns out, fitting the same model
in its state space form results in the same fit (see Table~\ref{LogParm}).  
\begin{table*}[t]
\begin{center}
\caption{SSM-Based Parameter Estimates of the Parametric Logistic Model}
\label{LogParm}
\begin{tabular}{@{}c c c c c @{}}
\hline
$constant$ & $scale$ &  $\phi$ & $\rho$  & $\sigma^{2}_{\epsilon}$ \\
\hline
3.605 & 1.844 & 1.398 & 0.104 & 0.0003 \\
\hline
\end{tabular}
\end{center}
\end{table*}
Note that, in the SSM formulation
only $\phi, \rho$ and $\sigma^{2}_{\epsilon}$ are estimated by the maximum likelihood.  The
other two parameters, $constant$ and $scale$, are obtained as a result of Kalman smoothing.  This is because,
Kalman smoothing provides the estimates of the two-dimensional latent state $\boldsymbol{\alpha}_{t}$ at
all times and $scale$ is the second element of $\boldsymbol{\alpha}_{t}$.  Furthermore, since the first element
of $\boldsymbol{\alpha}_{t}$ is $f(t)$, once $scale$, $\phi$, and $\rho$ are known, $constant$
can be derived from the estimate of $f(t)$ at some convenient $t$, say at $t = 0$.  Finally, Figure~\ref{logCurve}
shows the estimated Logistic fit (which matches the fit in \cite{bioGrow1}) with point-wise 95\% confidence band, 
which is also provided by the Kalman smoother.
\subsection{SSM Formulation of a Semiparametric Growth Curve}
\label{form2}
For a variety of reasons the parametric model described in Equation~\ref{eq1} rarely applies
perfectly in practice.  Instead, a less restrictive model of the following type is often more realistic:
\begin{equation}
y_{t} =  \mu_{t}+ \epsilon_{t} \label{eq2}
\end{equation}
where the latent curve, $\mu_{t}$, is suspected to be close to a parametric, monotone curve like 
$f(\boldsymbol{\theta}, t)$
but not necessarily exactly equal to it.   Adding a suitable disturbance term, $\boldsymbol{\eta}_{t}$, in the state
equation of Equation~\ref{BasicSSM}, provides a way to obtain such a semiparametric curve: 
\begin{eqnarray}
y_{t} & =  & [1 \; 0]  \; \boldsymbol{\alpha}_{t} + \epsilon_{t}  \; \;\;\quad\quad \quad\text{Observation Equation} \nonumber \\
\boldsymbol{\alpha}_{t+h} & = & \mathbf{T}_{t}^{t+h} \;\boldsymbol{\alpha}_{t}  + \boldsymbol{\eta}_{t+h}   
            \;\;  \;\; \quad \text{State Equation}  \label{semipara}\\
\boldsymbol{\alpha}_{0} & = &  \boldsymbol{\delta} \;\;\quad \quad \quad \quad\quad \quad\quad\quad
\text{Diffuse Initial Condition} \nonumber
\end{eqnarray}
Here the state disturbances, $\boldsymbol{\eta}_{t}$, are a sequence of independent, zero-mean, two-dimensional 
Gaussian vectors with appropriate choice of covariance $\mathbf{Q}_{t}$.   In this model the elements of
the latent state vector, $\boldsymbol{\alpha}_{t}$, don't have the same interpretation as in Equation~\ref{BasicSSM}.
Here $\boldsymbol{\alpha}_{t}[1] = \mu_{t}$, and $\boldsymbol{\alpha}_{t}[2]$ is no longer interpretable
as a simple time-invariant $scale$.

\begin{figure}[t]
\begin{center}
\includegraphics[scale=0.50]{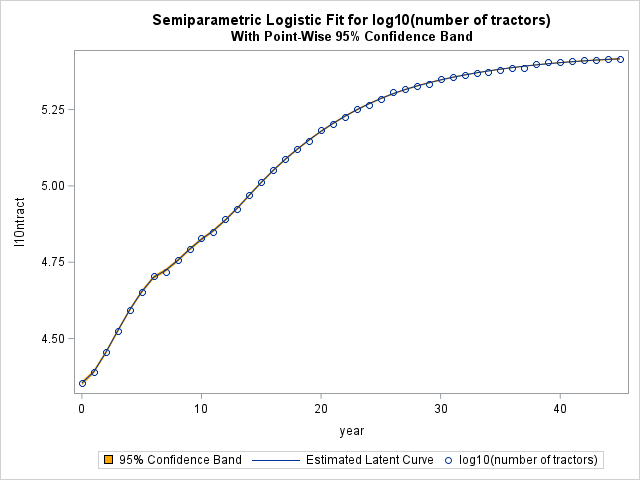}
\caption{Semiparametric Logistic Fit Based on the SSM Formulation}
\label{logSemiCurve}
\end{center}
\end{figure}

The state space formulations in Equation~\ref{BasicSSM} and Equation~\ref{semipara} 
are based on the treatment of semiparametric growth-curve modeling described in Section 3 of \cite{ans1}.  
There they describe how to obtain $\mathbf{Q}_{t}$ for different choices of $g(\boldsymbol{\theta}, t)$ so that the 
estimated latent curve, $\mu_{t}$, in Equation~\ref{eq2} satisfies a penalized least-squares criteria.  
However, except for
some simple choices of $g(\boldsymbol{\theta}, t)$, obtaining a closed form expression
for $\mathbf{Q}_{t}$ is complicated.  Instead, in this article we suggest always using the following greatly 
simplified form of $\mathbf{Q}_{t}$:
\begin{equation}
 \mathbf{Q}_{t}^{t+h} = \sigma^{2}_{\eta} \;
\begin{bmatrix}
\Delta^{3}/3 &\Delta^{2}/2\\
\Delta^{2}/2 & \Delta
\end{bmatrix}
 \label{specialQ}
\end{equation}
where $\sigma^{2}_{\eta}$ is a positive scaling factor and $\Delta = |g(\boldsymbol{\theta}, t+h) - g(\boldsymbol{\theta}, t)|$.
We have found that this choice of $\mathbf{Q}_{t}$ works well for a wide variety of $g(\boldsymbol{\theta}, t)$ used 
in practice.  Note that,
\begin{itemize}
\item The transition matrix, $\mathbf{T}_{t}^{t+h}$, used in the
formulations in Equation~\ref{BasicSSM} and Equation~\ref{semipara}, is exactly the same as in
 Section 3 of \cite{ans1}.
\item The choice of $\mathbf{Q}_{t}$ described in Equation~\ref{specialQ} is exact for the linear 
case, $g(\boldsymbol{\theta}, t) = t$.  
In the linear case, the resulting
estimate of $\mu_{t}$ is the well-known cubic smoothing spline fit.
Our recommendation of this form of $\mathbf{Q}_{t}$ even for a nonlinear $g(\boldsymbol{\theta}, t)$
is based on it being nearly optimal for a piecewise linear approximation of $g(\boldsymbol{\theta}, t)$.
\item For the exponential case, $g(\boldsymbol{\rho}, t) = \exp(-\rho \;t)$, a closed
form expression for $\mathbf{Q}_{t}$ is described in \cite{ans1} and an equivalent form of this
model is discussed in \cite{deJong}.   For this exponential case, we compared the growth/decay curves
estimated using the exact form of $\mathbf{Q}_{t}$ versus the simplified form in Equation~\ref{specialQ}
in many real and simulated data cases.
In all cases the estimated curves by the exact and approximate methods matched very well.  
In particular, for the chlorine content
data example discussed in both \cite{ans1}  and \cite{deJong}, the estimated curves were
essentially identical.
\item When $\sigma^{2}_{\eta}$, the scaling factor of $\mathbf{Q}_{t}$, is either set to zero or if 
its estimate is nearly zero, the fully parametric case
emerges.  Therefore, the degree of conformity of 
the data to the fully parametric form can be assessed by the closeness of the estimate of $\sigma^{2}_{\eta}$ to zero.
\end{itemize}
Continuing with the \emph{greek\_tractors} data set, Table~\ref{LogSemi} shows the
parameter estimates of the semiparametric Logistic fit.
Since the estimate of 
the scaling factor, $\sigma^{2}_{\eta} = 99.77$, is far from zero, the
estimated curve, $\mu_{t}$, departs from the parametric Logistic form and the parameters
$\phi$ and $\rho$ don't have their traditional meaning.  
Nevertheless, the estimate of $\rho = 0.109$ is close to $\rho$ in the parametric form
(see Table~\ref{LogParm}).
Figure~\ref{logSemiCurve}
shows the estimated curve, $\mu_{t}$, with point-wise 95\% confidence band.

In the semiparametric case, a closed form expression for $\mu_{t}$, like in Equation~\ref{logEq},
 is not possible.  You can, however, evaluate $\mu_{t}$ at any point
$t$ by including an (artificial) observation at that point with a missing response value.
The SSM formulation handles missing response values with ease and the Kalman
smoother produces appropriate estimate of $\mu_{t}$ at such points.  
\begin{table*}[t]
\begin{center}
\caption{SSM-Based Parameter Estimates of the Semiparametric Logistic Model}
\label{LogSemi}
\begin{tabular}{@{}c c c c @{}}
\hline
$\phi$ & $\rho$ &$\sigma^{2}_{\eta}$  & $\sigma^{2}_{\epsilon}$ \\
\hline
 0.324 & 0.109 & 99.77& 0.000 \\
\hline
\end{tabular}
\end{center}
\end{table*}

\subsection{FME Model With Growth Curve As a Functional Fixed-Effect}
\label{form3}
Let $\{y_{t}^{i}, i = 1, 2,\cdots, K\}$ denote longitudinal growth (or decay) measurements on $K$ independent 
experimental units, which share similar environment.  For example, they
might denote growth patterns of 
puppies from the same litter, or
the number of units sold over time of a new product in different (but similar) regions. 
 The following FME model can be used to describe
such data:
\begin{equation}
y_{t}^{i} =  \mu_{t}+ \omega_{t}^{i} + \epsilon_{i,t} \label{fme1}
\end{equation}
Here, $\mu_{t}$, the mean curve, is a parametric or semiparametric growth curve, 
$\{ \omega_{t}^{i},  i = 1, 2,\cdots, K\}$ are deviation curves of the individual units from the mean curve,  
and $\{ \epsilon_{i, t},  i = 1, 2,\cdots, K\}$ are independent, $\text{N}(0,  \sigma^{2}_{\epsilon})$ errors.
In the FME literature, $\mu_{t}$ is called the functional fixed-effect and $\omega_{t}^{i}$
are called functional-random-effects.  The deviation curves are often modeled as zero-mean, 
random-walks, or more elaborately as zero-mean, autoregressive processes (or something similar).
For the illustrations in this article we will use zero-mean,  random walks $\{ \omega_{t}^{i}\}$, which are defined as
\begin{eqnarray}
\omega_{t+h} & = & \omega_{t} + \xi_{t+h} \nonumber \\
\omega_{0} & = & 0 \nonumber
\end{eqnarray} 
where the disturbances $\xi_{t+h}$ are independent  $\text{N}(0,  \sigma^{2}_{dev} h)$ variables
($\sigma^{2}_{dev}$ is a positive scaling parameter).  
The FME model described in Equation~\ref{fme1} and much more general FME models, e.g., those with more
general deviation models and with regression effects, can be easily formulated as SSMs.  For now, we
just describe the SSM corresponding to  Equation~\ref{fme1} with $\{\omega_{t}^{i}\}$ as zero-mean,  random walks:
\begin{eqnarray}
y_{t}^{i} & =  &  \mathbf{Z}^{i} \; \boldsymbol{\alpha}_{t} + \epsilon_{i,t}  \quad\quad \quad\text{Observation Equation} \nonumber \\
\boldsymbol{\alpha}_{t+h} & = & \mathbf{T}_{t}^{t+h} \;\boldsymbol{\alpha}_{t}  + \boldsymbol{\eta}_{t+h}   
            \quad \text{State Equation}  \label{StochasticSSM}\\
\boldsymbol{\alpha}_{0} & = &   \begin{bmatrix} \boldsymbol{\delta} \\
0 
\end{bmatrix} \;\quad \quad\quad \quad\quad\quad
\text{Partially Diffuse Initial Condition} \nonumber
\end{eqnarray}
where the latent state vectors, $\boldsymbol{\alpha}_{t}$, are of dimension $(2+K)$, $\mathbf{Z}^{i}$ is a 
$(2+K)$-dimensional row vector with $\mathbf{Z}^{i}[1] = 1, \;\mathbf{Z}^{i}[2+i] = 1$ and all other elements are equal 
to zero, $(2+K)$-dimensional transition matrix $\mathbf{T}_{t}^{t+h}$ is block-diagonal with the first 2-dimensional
block same as that in Equation~\ref{semipara} and the second block is the $K$-dimensional
identity matrix, the covariance of $\boldsymbol{\eta}_{t+h}$, $\mathbf{Q}_{t}^{t+h}$ is also block
diagonal with the first block same as that in Equation~\ref{specialQ} and the 
second block is the $K$-dimensional identity matrix multiplied by $\sigma^{2}_{dev} h$.

\begin{figure}[t]
\begin{center}
\includegraphics[scale=0.50]{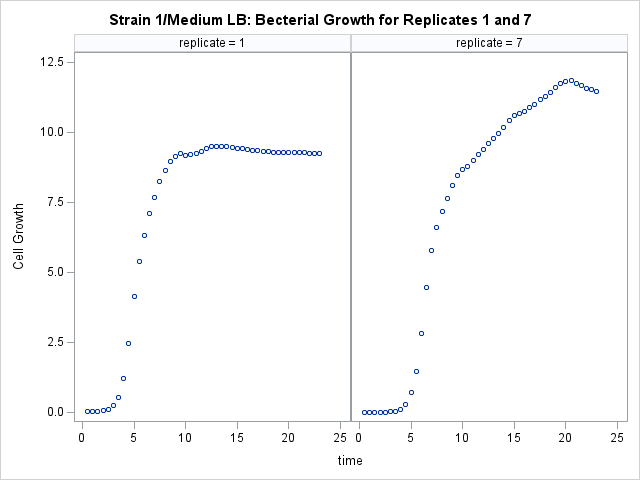}
\caption{Bacterial growth for Strain 1 in medium LB.}
\label{scatter1}
\end{center}
\end{figure}

\begin{figure}[t]
\begin{center}
\includegraphics[scale=0.50]{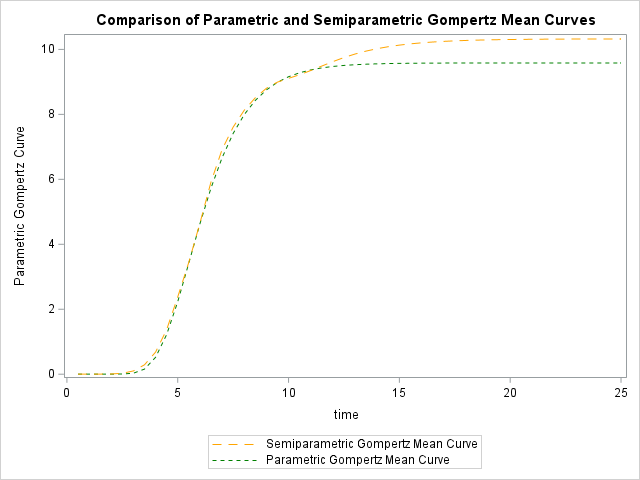}
\caption{Comparison of mean-curve estimates of parametric and semiparametric Gompertz models}
\label{PgNg}
\end{center}
\end{figure}

\begin{figure}[t]
\begin{center}
\includegraphics[scale=0.50]{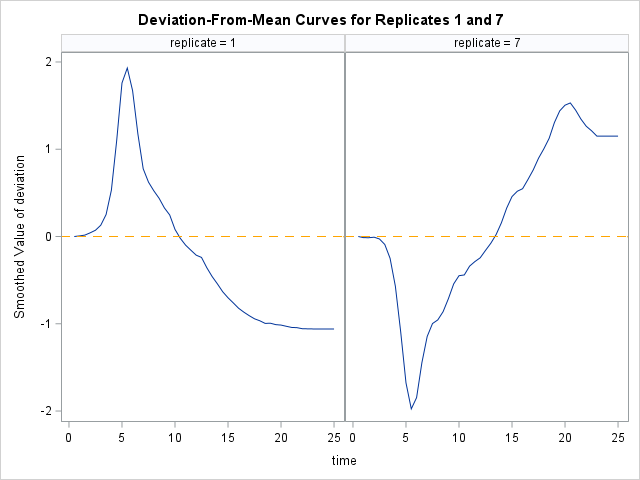}
\caption{Deviation curves for replicates 1 and 7}
\label{dev17}
\end{center}
\end{figure}

As a quick illustration of FME model based data analysis, we analyze a small portion of 
the data used in \cite{jap}.
Slightly more elaborate analysis of these data is done in
Section~\ref{example}.  Here we consider 12 independent replicate curves generated by 
monitoring the growth of a bacterium strain in a nutrient rich medium (strain 1, medium LB).
The measurements are taken every 30 minutes during a period of 23 hours.
Figure~\ref{scatter1} shows the growth patterns for replicates 1 and 7.  It shows
that both the replicates loosely follow the usual sigmoidal shape, however,
they do show noticeable differences in their growth patterns.
We fit two FME models, one with the parametric and the other with the semiparamtric Gompertz curve 
as a mean-curve.  Table~\ref{GFme} shows the parameter estimates and the BIC
information criteria for these two models.  On the basis of BIC, the FME model based
on the semiparametric Gompertz curve as a mean-curve is preferred.
\begin{table*}[t]
\begin{center}
\caption{FME Model Fitting for Parametric and Semiparametric Gompertz Curves}
\label{GFme}
\begin{tabular}{@{}c c c c c c c c c@{}}
\hline
Parametric & $Constant$ & $Scale$ & $\phi$ &  $\rho$ &   $\sigma^{2}_{\eta}$ & 
$\sigma^{2}_{dev}$ & $\sigma^{2}_{\epsilon}$ & BIC \\
\hline
Yes & 0.003 &  9.58     &  46.97  & 0.69  &  0  &  0.036  & 0.00015  & -299.95    \\
No & NA         &  NA           &  20.91 &  0.46  &  102.03  &  0.034  &  0.00014 & -312.63 \\
\hline
\end{tabular}
\end{center}
\begin{tablenotes}
\item  "NA" in a parameter column indicates that the parameter is not applicable for the 
model type in the corresponding row.
\end{tablenotes}
\end{table*}
Figure~\ref{PgNg} shows the mean-curve estimates based on the two models.
It shows that the saturation level implied by the two models is quite
different.  Saturation level is often an important parameter to estimate.
The estimated deviation curves highlight the difference between the
replicate curves and the mean-curve.  Continuing the
exploration of the output from the preferred model, Figure~\ref{dev17}
shows such deviation curves for the replicates 1 and 7.  This type of
examination can alert the researchers to outlying replicates.
\section{Illustration: Comparing Bacterial Growth in Different Settings}
\label{example}
In \cite{jap}, correlation between genome reduction and bacterial growth is studied 
(a link to access their data is also provided in the article).
In their study, E. Coli cultures are grown in three settings of growth mediums and at
36 different levels of gene deletion.  For each combination of growth medium and
gene deletion level, multiple replicates of E. Coli  cultures are grown.
Their growth is measured at 30 minute or one hour intervals for different lengths
of times.  The three growth mediums are M63, MAA, and LB, which indicate poor, supplementary,
and rich growth conditions, respectively.  The 36 gene deletion levels are labeled as
strains 1 to 36, which indicate increasing severity of gene deletion.  
In our illustration 
we will analyze the growth patterns for strains 1 and 36, grown at all the three mediums,
M63, MAA, and LB (that is, total of 6 strain-medium combinations).
Except in the case of (strain 36, medium M63) combination, which has 24 replicates,
all other combinations have 12 replicates.  Even though our methodology can
easily handle different number of replicates for different combinations, we excluded
the last 12 replicates for the (strain 36, medium M63) combination.  Additionally, two minor
adjustments to the data were made for easier viewing of the results: 
\begin{itemize} 
\item arificial observations with missing response values were added at 30 minute intervals
to ensure that all the replicates have measurements on the same, uniform, time grid.  
This ensured that all the component estimates can be plotted in a similar way.
\item all the growth measurements were multiplied by 10.0, which helped in some
graphical displays.
\end{itemize}

\begin{figure}[t]
\begin{center}
\includegraphics[scale=0.50]{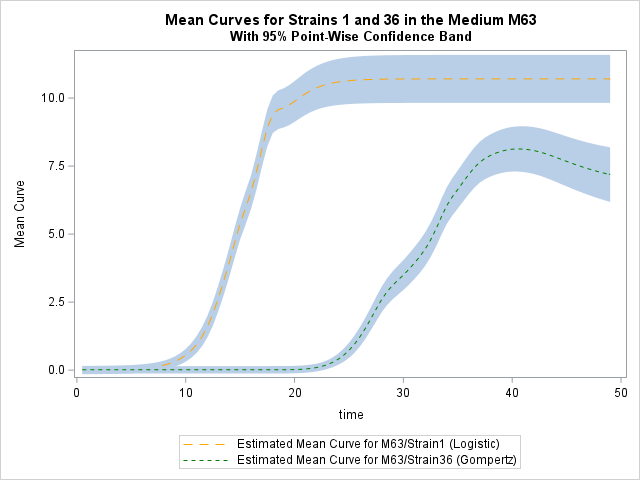}
\caption{Change in the mean-curves from the least to most gene deletion for the medium M63}
\label{m63}
\end{center}
\end{figure}

\begin{figure}[t]
\begin{center}
\includegraphics[scale=0.50]{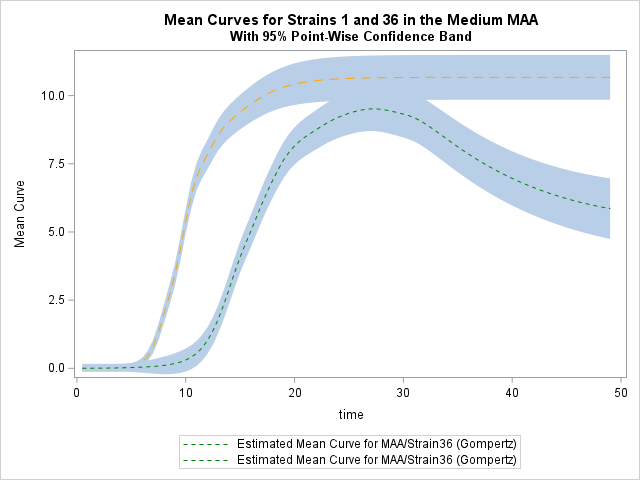}
\caption{Change in the mean-curves from the least to most gene deletion for the medium MAA}
\label{maa}
\end{center}
\end{figure}

\begin{figure}[t]
\begin{center}
\includegraphics[scale=0.50]{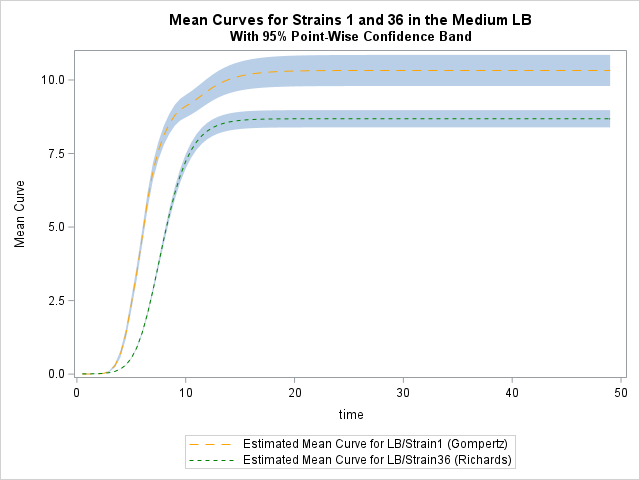}
\caption{Change in the mean-curves from the least to most gene deletion for the medium LB}
\label{lb}
\end{center}
\end{figure}

We began our analysis by trying four FME models where the mean-curves
are modeled as semiparametric Linear, Logistic, Gompertz, and Richards 
curves, respectively, for each of the six strain/medium combinations.  The model
with best BIC criteria was chosen for each combination.  Table~\ref{FParm} shows
the chosen models and their parameter estimates.
\begin{table*}[t]
\begin{center}
\begin{threeparttable}[b]
\caption{Parameter Estimates For The Chosen Models}
\label{FParm}
\begin{tabular}{@{}c l l r  r c r r r@{}}
\hline
Strain & Medium &  Prior & $\phi$ &  $\rho$ &  $\nu$ &  $\sigma^{2}_{\eta}$ & $\sigma^{2}_{dev}$ & $\sigma^{2}_{\epsilon}$ \\
\hline
1 & M63 &  Logistic &  9517.73  & 0.559  & NA &  244.54 &  0.073  & 0.00000 \\
 & MAA &  Gompertz  & 39.58 &  0.372 &  NA &  145.63 &  0.065 &  0.00000 \\
 & LB &  Gompertz  &  20.91 &  0.463  & NA &  102.03  &  0.034  &  0.00014 \\
\hline
\hline
36 & M63 & Gompertz &  233.67 &  0.172  & NA   &  366.96 & 0.059 & 0.00000  \\
 & MAA & Gompertz     &    7.90 & 0.118 & NA & 793.04 & 0.044 & 0.00000  \\
 & LB &   Richards  &  71.69 &   0.674  &  0.445  &  0.00  &  0.019  &  0.00004 \\
\hline
\end{tabular}
\begin{tablenotes}
\item  "NA" in a parameter column indicates that the parameter is not applicable for the 
model type in the corresponding row.
\end{tablenotes}
\end{threeparttable}
\end{center}
\end{table*}
As you can see,  for all the combinations a traditional growth-curve model is chosen for the mean-curve.  
For the (strain 36, medium LB) combination, because the estimated scaling factor 
$\sigma^{2}_{\eta}$ is nearly zero, the mean curve turns out be a parametric Richards curve.
The change in the mean curve between the least gene deletion (strain 1) and the most severe
gene deletion (strain 36) for different mediums is shown in Figure~\ref{m63} for M63,
in Figure~\ref{maa} for MAA, and in Figure~\ref{lb} for LB.
A few things are apparent from these plots:
\begin{itemize}
\item  For each medium, the mean curves shift to the right 
and the saturation level (or the maximum level) goes down as gene deletion increases. 
\item In medium LB, which is the most nutrient rich, the mean curves follow the usual sigmoidal form
for both strain 1 and strain 36 (see Figure~\ref{lb}).
\item In mediums M63 and MAA (Figure~\ref{m63} and Figure~\ref{maa}), which are nutritionally poorer, the mean curves for Strain 1 follow
the sigmoidal form.  However, for Strain 36 the mean level departs significantly from the sigmoidal form and 
the mean level goes down after reaching its peak.  That is, in these two mediums the gene deletion
causes significant changes in the growth pattern.
\end{itemize} 
Since all but one of these mean curves are semiparametric, the usual meanings of 
growth curve model parameters don't apply.  So their peak growth rates, time of peak growth, etc.,
cannot be inferred from these parameters.  Instead, these must be deduced from the estimated
mean curves themselves.  Since we have evaluated these mean curves, say $\hat{\mu}_{t}$, on a uniform 
time grid (of 30 minute interval), we can approximate their growth rate curves, say $\hat{\lambda}_{t}$, 
 by differencing: that is, $\hat{\lambda}_{t} = \hat{\mu}_{t} - \hat{\mu}_{t-1}$.
Table~\ref{FInfo} shows some summary information derived from these growth rate curves.  
  It shows that for all three mediums
the peak growth rate for strain 36 is considerably lower than the peak growth rate for strain 1,
moreover, in each case the peak growth rate is attained at a later time.
\begin{table*}[t]
\begin{center}
\begin{threeparttable}[b]
\caption{Maximum Growth Rates of the Mean Curves (Based on $\hat{\lambda}_{t}$)}
\label{FInfo}
\begin{tabular}{@{}l c r r @{}}
\hline
 Medium  & Strain   &  Max Growth Rate & Attained at Time\\
\hline
 M63 & 1  &  0.84 &   17.0   \\
 M63 & 36  & 0.45  &   33.5 \\
\hline
 MAA &  1 & 1.03 &10.0    \\
MAA & 36   & 0.59 &      14.5\\
\hline
 LB     &1  &  1.27 &   6.5   \\
 LB     &  36  & 0.88   &   8.0  \\
\hline
\end{tabular}
\end{threeparttable}
\end{center}
\end{table*}
The SSM framework can provide a lot more insight into the data.  For example, it is easy to identify 
outlying observations and structural breaks in the components.  Additionally, you can estimate difference 
between two component curves (e.g., the difference in the mean curves between two strain/medium 
combinations) and obtain confidence bands for such differences.  This is useful for testing
hypotheses about statistically significant differenses between such curves.  For this illustration
also we could have continued to explore the data further but we will stop here.  Of course,
for the overall study, we did not analyze data for strains 2 to 35 at all.  These strains could be
analyzed in a similar fashion.
\section{Conclusion}
\label{final}
We have described a flexible SSM-based framework for semiparametric modeling of growth curve
data that are generated by hierarchical, longitudinal studies.  Our modeling approach is based
on a simplification of growth curve modeling approach that is described in \cite{ans1}. 
We hope our treatment of their approach is more accessible and easier to adopt.  
The SSM-based framework described here can be easily extended in many directions.
For example, the FME model in Equation~\ref{fme1} can contain regression effects, or
additional effects that modify the mean curve to account for
hormetic effects.  

\section[Disclaim]{Disclaimer}
The views and opinions expressed in this article are solely those of the author 
and do not necessarily reflect the official policy or position of the author's employer. 
The accompanying code is provided "as is," without any warranties, express or implied, including but not 
limited to the implied warranties of merchantability and fitness for a particular purpose. 
The author and the employer shall not be liable for any damages arising from the use of the code.
\appendix
\section[SSMFramework]{SSM Framework and Notation}
\label{SSMFramework}
 All the SSMs discussed in this article are special cases of
the following form:
\begin{eqnarray}
\mathbf{y}_{t} & = & \mathbf{Z}_{t}  \boldsymbol{\alpha}_{t} + 
                                      \mathbf{X}_{t}  \boldsymbol{\beta} + \boldsymbol{\epsilon}_{t}  
                                       \quad \text{Observation Equation} \nonumber \\
 \boldsymbol{\alpha}_{t+1}  & = & \mathbf{T}_{t}^{t+1}  \boldsymbol{\alpha}_{t} +
                                                              \boldsymbol{\eta}_{t+1}  
                                                              \quad \quad\text{State Equation}  \label{basic} \\
 \boldsymbol{\alpha}_{0}  & = &    \boldsymbol{\eta}_{0} \quad \quad \quad \quad \quad
                                               \quad \; \; \; \;\;\text{Partially Diffuse Initial Condition} \nonumber 
\end{eqnarray}
\begin{itemize}
\item $\mathbf{y}_{t} , t = 1, 2, \cdots$ is a sequence of response vectors.  The index $t$
traverses the set $(\tau_{1} < \tau_{2}  <  \cdots)$, which are the actual time points at which the responses are
measured.  The data are longitudinal and the distances between the successive time points 
need not be the same, that is, $h_{1} =  (\tau_{2} -  \tau_{1})$
and $h_{2} =  (\tau_{3} -  \tau_{2})$ need not be the same.  The number
of responses at different times, i.e., the dimension of $\mathbf{y}_{t}$ at different times, need not be
the same and, some or all elements of $\mathbf{y}_{t}$ can be missing.  In fact, missing
measurements indicate that their values are to be predicted using the remaining
observed data.
\item The observation equation expresses the response vector as 
a sum of three terms: $\mathbf{Z}_{t}  \boldsymbol{\alpha}_{t}$ denotes
the contribution of the state vector $\boldsymbol{\alpha}_{t}$,
$\mathbf{X}_{t}  \boldsymbol{\beta}$ denotes
the contribution of the regression vector $\boldsymbol{\beta}$,
and $\boldsymbol{\epsilon}_{t}$ is a zero-mean, Gaussian noise vector with
diagonal covariance matrix.  The dimension of the state vector, $\boldsymbol{\alpha}_{t}$, does not change
with time.  The design matrices $\mathbf{Z}_{t}$ and $\mathbf{X}_{t}$ are of compatible dimensions.  
\item According to the state equation,  $\boldsymbol{\alpha}_{t+1}$, the state at time $(t+1)$,
is a linear transfromation of the previous state, $\boldsymbol{\alpha}_{t}$, plus a random
disturbance, $\boldsymbol{\eta}_{t+1}$, which is a zero-mean, Gaussian vector with
covariance $\mathbf{Q}_{t}^{t+1}$ that need not be diagonal.  
The elements of the state transition matrix $\mathbf{T}_{t}^{t+1}$ and the
disturbance covariance $\mathbf{Q}_{t}^{t+1}$
are known and can depend on information available at times $t$ and $t+1$.
\item The initial state, $\boldsymbol{\alpha}_{0}$, is assumed to be a Gaussian vector with
known mean, and covariance $\mathbf{Q}_{0}$.  In many cases, no prior informarion about
some elements of $\boldsymbol{\alpha}_{0}$ is available.   In such cases, their variances 
are taken to be infinite and these elements are called diffuse.
\item The noise vectors in the observation and state equations, $\boldsymbol{\epsilon}_{t}$,
$\boldsymbol{\eta}_{t}$, and the initial condition $\boldsymbol{\alpha}_{0}$, are assumed
to be mutually independent.
\item The elements of system matrices $\mathbf{Z}_{t}, \text{Cov}(\boldsymbol{\epsilon}_{t}), 
\mathbf{T}_{t}^{t+1}, \mathbf{Q}_{t}^{t+1}$, and $\mathbf{Q}_{0}$
are assumed to be completely known, or some of them can be functions of a small set of unknown 
parameters (to be estimated from the data). 
\end{itemize}
The latent vector $\boldsymbol{\alpha}_{t}$ 
can often be partioned into meaningful subblocks (with corresponding blocking
of the design matrix $\mathbf{Z}_{t}$).  In these cases the
observation equation in Equation~\ref{basic} gets the following form:
\[
\mathbf{y}_{t} = \boldsymbol{\mu}_{t} + \boldsymbol{\gamma}_{t} + \cdots + 
                              \mathbf{X}_{t}  \boldsymbol{\beta} + \boldsymbol{\epsilon}_{t}  
\]
where the terms $\boldsymbol{\mu}_{t}$, $ \boldsymbol{\gamma}_{t}, \cdots$
might represent a time-varying mean-level, a seasonal pattern, and so on.
Such linear combinations of the state subblocks are called components.  
When the data from a longitudinal study is assumed to follow an SSM, 
the data analyis is greately helped by the well-known
(diffuse) Kalman filter and (diffuse) Kalman smoother algorithms.  Chapters 4, 5, 6, and 7 of \cite{dk}
explain how these algorithms provide the following:
\begin{itemize}
\item Maximum likelihood estimates of the unknown model parameters that are obtained by maximizing
the marginal likelihood.
\item A variety of diagnostic measures for model evaluation.
\item Full-sample estimates of the latent vectors 
$\boldsymbol{\alpha}_{t}$, $\boldsymbol{\beta}$, and the model components 
such as $\boldsymbol{\mu}_{t}, \boldsymbol{\gamma}_{t}, \cdots$, at all time points.
The full-sample estimates are also called the smoothed estimates in the SSM literature.
\item Full-sample predictions of all missing response values.
\end{itemize}

\bibliography{growthBib}

\begin{thebibliography}{14}
\providecommand{\natexlab}[1]{#1}
\providecommand{\url}[1]{\texttt{#1}}
\expandafter\ifx\csname urlstyle\endcsname\relax
  \providecommand{\doi}[1]{doi: #1}\else
  \providecommand{\doi}{doi: \begingroup \urlstyle{rm}\Url}\fi

\bibitem[Ansley et~al.(1993)Ansley, Kohn, and Wong]{ans1}
C.~F. Ansley, R.~Kohn, and C.~M. Wong.
\newblock Nonparametric spline regression with prior information.
\newblock \emph{Biometrika}, 80\penalty0 (1):\penalty0 75--88, 1993.

\bibitem[Blozis and Harring(2016)]{blozis}
S.~A. Blozis and J.~R. Harring.
\newblock On the estimation of nonlinear mixed-effects models and latent curve
  models for longitudinal data.
\newblock \emph{Structural Equation Modeling}, 23\penalty0 (6):\penalty0
  904--920, 2016.

\bibitem[Chan et~al.(2021)Chan, Tsui, Wei, Zhang, and Deng]{MonoSpline2}
V.~Chan, K.~Tsui, Y.~Wei, Z.~Zhang, and X.~Deng.
\newblock Efficient estimation of smoothing spline with exact shape
  constraints.
\newblock \emph{Statistical Theory and Related Fields}, 5\penalty0
  (1):\penalty0 55--69, 2021.

\bibitem[De~Jong and Mazzi(2001)]{deJong}
P.~De~Jong and S.~Mazzi.
\newblock Modeling and smoothing unequally spaced sequence data.
\newblock \emph{Statistical Inference for Stochastic Processes}, 4\penalty0
  (1):\penalty0 53--71, 2001.

\bibitem[Durbin and Koopman(2012)]{dk}
J.~Durbin and S.~J. Koopman.
\newblock \emph{Time Series Analysis by State Space Methods, 2nd ed.}
\newblock Oxford University Press, Oxford, 2012.

\bibitem[Fine et~al.(2019)Fine, Won~Suk, and Grimm]{grimm}
L.F. Fine, H.~Won~Suk, and K.~J. Grimm.
\newblock An examination of a functional mixed-effects modeling approach to the
  analysis of longitudinal data.
\newblock \emph{Multivariate Behavioral Research}, 54\penalty0 (4):\penalty0
  475--491, 2019.

\bibitem[Garre et~al.(2023)Garre, Koomen, den Besten, and Zwietering]{bioGrow1}
A.~Garre, J.~Koomen, J.~den Besten, and H.M.W. Zwietering.
\newblock Modeling population growth in r with the biogrowth package.
\newblock \emph{Journal of Statistical Software}, 107\penalty0 (1), 2023.

\bibitem[Guo(2002)]{Guo1}
W.~Guo.
\newblock Functional mixed effects models.
\newblock \emph{BIOMETRICS}, 58:\penalty0 121--128, 2002.

\bibitem[Harvey and Kattuman(2020)]{Covid1}
A.~Harvey and P.~Kattuman.
\newblock Time series models based on growth curves with applications to
  forecasting coronavirus.
\newblock 1, 2020.
\newblock \doi{https://doi.org/10.1162/99608f92.828f40de}.

\bibitem[Kurokawa et~al.(2016)Kurokawa, Seno, Matsuda, and Ying]{jap}
M.~Kurokawa, S.~Seno, H.~Matsuda, and B.~Ying.
\newblock Correlation between genome reduction and bacterial growth.
\newblock \emph{DNA Research}, 23\penalty0 (6):\penalty0 425--441, 2016.

\bibitem[Ramsay(1988)]{MonoSpline1}
J.O. Ramsay.
\newblock Monotone regression splines in action.
\newblock \emph{Statistical Science}, 3:\penalty0 517--525, 1988.

\bibitem[SAS()]{cssm1}
\emph{The CSSM Procedure, SAS/Econometrics User's Guide}.
\newblock SAS Institute Inc., Cary, NC.
\newblock
  \url{https://go.documentation.sas.com/doc/en/pgmsascdc/v\_059/casecon/casecon\_cssm\_toc.htm}.

\bibitem[Selukar(2015)]{selu_r:15}
R.~S. Selukar.
\newblock Functional modeling of longitudinal data with the ssm procedure.
\newblock In \emph{Proceedings of the SAS Global Forum 2015 Conference}, Cary,
  NC, 2015. SAS Institute Inc.
\newblock
  \url{https://support.sas.com/resources/papers/proceedings15/SAS1580-2015.pdf}.

\bibitem[Zwietering et~al.(1990)Zwietering, Jongenburger, Rombouts, and Van
  'T~Riet]{zw}
M.~H. Zwietering, I.~Jongenburger, F.M. Rombouts, and K.~Van 'T~Riet.
\newblock Modeling of the bacterial growth curve.
\newblock \emph{Applied and Environmental Microbiology}, 1:\penalty0
  1875--1881, 1990.

\end{thebibliography}
\end{document}